\documentclass[conference]{IEEEtran}
\ifCLASSINFOpdf
   \usepackage[pdftex]{graphicx}
   \DeclareGraphicsExtensions{.pdf,.jpeg,.png,.eps}
\else
\fi
\usepackage{amsmath}
\usepackage{amssymb}
\usepackage{epstopdf}

\begin{document}
\pagestyle{empty}
\thispagestyle{empty}
\title{A Tractable Approach to Dynamic Network Dimensioning Based on the Best-cell Configuration}

\author{\IEEEauthorblockN{Yanyan Wu}\\
\IEEEauthorblockA{
Xi'an Jiatong-Liverpool University\\
215123 Suzhou, P R China\\
Email: yanyan.wu@xjtlu.edu.cn
}

}


%


\maketitle

\begin{abstract}

Spatial distributions of other cell interference (OCIF) and interference to own-cell power ratio (IOPR) with reference to the distance between a mobile and its serving base station (BS) are modeled for the down-link reception of cellular systems based on the best-cell configuration instead of the nearest-cell configuration. This enables a more realistic evaluation of two competing objectives in network dimensioning: coverage and rate capacity. More outcomes useful for dynamic network dimensioning are also derived, including maximum BS transmission power per cell size and the cell density required for an adequate coverage of a given traffic density.

\end{abstract}


%

\section{Introduction}
More and more base stations have been deployed all over the world to meet an ever-increasing demand for more date rate per unit area. Base station power consumption, being the major part of the energy consumption of a wireless network, becomes a target for energy efficiency optimisation \cite{Auer2011}. On one hand, cells are getting smaller and smaller in order to provide more capacity. On the other, dynamic dimensioning that caters for varying traffic both temporally and geographically appears to be promising for energy saving \cite{niu2011tango}\cite{eunsung2013}.

 Dynamic dimensioning relies on a tractable approach to coverage and capacity that is based on the analytical expression of spatial distribution of SIR with reference to the distance between a mobile and its serving BS. Two representative modelings were found in \cite{Kelif2009CellBreathing} for a grid network model and in \cite{andrews2011} for a Poisson point process (PPP) distributed network model. The latter was extended to optimal BS density design and dynamic BS sleeping in \cite{Cao}. Both modelings assumed that users were served by the nearest BS regardless of shadowing effect, however in cellular operation, the cell search procedure often allows a mobile to attach to a BS from which it receives the strongest signal, thus the best cell is not necessarily the nearest one due to channel shadowing effect.  With the best-cell configuration, the coverage area of a BS may go well beyond the physical boundary of a cell, and the statistical distribution of SIR is different from that of the nearest cell configuration owing to the truncation effect of cell selection. For this reason, unrealistic outage probabilities were drawn in \cite{Kelif2009Shadowing}. The best-cell configuration was considered in \cite{Dhillon2012} for PPP network with two questionable assumptions: i) SINR is greater than one when in coverage, ii) Rayleigh fading in stead of log-normal (LN) shadowing effect. The first assumption can not be true for a CDMA system with large processing gain. On the second, log-normal shadowing with a typical standard deviation between 8 dB and 12 dB models gradual changes over a large distance in cellular networks, therefore is more relevant to cell planning than Rayleigh fading that models multi-path effects (``microscopic" changes over a small distance comparable to the carrier wavelength).

These above reasons motivated the work in this paper.  Following an investigation of mobile distributions with reference to the distance to the serving BS, analytical modeling of OCIF and IOPR are formulated for the downlink reception with the best-cell configuration for a grid model network. The results are shown to be applicable to both CDMA and OFDMA based systems. Moreover maximum BS transmission power is derived per cell size for an interference limited system.

In section II, the system parameters are specified with reference to \cite{Kelif2009CellBreathing}.  Mobile distribution over the whole coverage area as a result of cell assignment by BS signal strength is investigated in Section III, following which the spatial distributions of both OCIF and IOPR with reference to the distance between a mobile and its serving BS are modeled in Section IV. These allow a more realistic evaluation of coverage and capacity in Section V, in which more outcomes for dynamic network dimensioning are also derived for an OFDMA system including a) maximum transmission power per cell size, b) cell density required for a given traffic density. Conclusions are given in section VI.

\section{System model}
  Consider a grid model homogenous cellular network, each BS is located at the center of a hexagonal cell with a regular distance of $2R_{c}$ between them. Each BS transmits the same power via an omni-directional antenna. Mobile users are evenly distributed in the whole area. Due to both path loss and shadowing effect in wireless propagation, it is anticipated that a mobile is less likely to attach to a BS with increasing distance, thus the number of users attached to a BS diminishes gradually with increasing distance between the users and the BS.

\begin{itemize}
\item $P_b=P_{cch}+\Sigma_{u}P_{b,u}$ is the total power transmitted by BS $b$, including power of common control channel $P_{cch}$ and all mobiles $P_{b,u}$;
\item $G_{b,u}$ denotes the wireless channel propagation gain from BS $b$ to the mobile $u$, following a combined path-loss and shadowing model \cite{Goldsmith}:
              \begin{align}
                    G_{b,u}= k_0(r_b/r_0)^{-\eta}e^{-a\xi_b}  \quad  (r_b>r_0)
            \end{align}
where $r_b$ is the distance between a mobile and its serving BS $b$, $\eta$ is the path-loss exponent, $k_0$ is radio frequency related attenuation at near field ($r_b<r_0$) and $a=ln10/10$. $k_0=-10$dB at $r_0=1$ m are assumed in this paper. $e^{-a\xi_b}$ follows a log-normal distribution whilst the shadowing factor $\xi_b$ has a probability density function (pdf) $p(\xi_b)=N(0,\sigma^{2})$. Assume also that the log-normal shadowing from different base stations are independent.
              \item $I_{int,u}=P_bG_{b,u}$ denotes the own-cell power received at mobile $u$;
               \item $I_{ext,u}=\sum_{i\neq b}P_iG_{i,u}$ denotes the OCIF from co-channel cells received at mobile $u$.
\end{itemize}


\subsection {CDMA system}

Assume perfect power control in CDMA system, and the SINR target $\gamma^{*}$ is the same for all users for a basic service. According to \cite{Kelif2009CellBreathing}, the transmission power of BS $b$ can be computed by
     \begin{align}\label{eqn:Pcdma}
                P_b=P_{cch}+\sum_u{P_{b,u}}=\frac{P_{cch}+\frac{\gamma^{*}\sigma^2_{n}}{1+\alpha\gamma^{*}}\sum_u{h_u}}{1-\frac{\gamma{*}}{1+\alpha\gamma{*}}\sum_u(\alpha+ f_u)}
     \end{align}
where
\begin{align}\label{eqn:fu}
f_u=\frac{I_{ext,u}}{I_{int,u}}=\frac{\sum_{i\neq b}G_{i,u}}{G_{b,u}}
\end{align}
\begin{align}\label{eqn:hu}
h_u=\frac{1}{G_{b,u}}=\frac{r_b^{\eta}e^{a\xi}}{k_0r_0^{\eta}}
\end{align}
and $\alpha$ is the orthogonality factor that accounts for the imperfection of orthogonality between signals within the same cell, $\sigma^2_{n}$ is the power of additive white Gaussian noise (AWGN).

Note $f_u$ which gives IOPR of a homogenous CDMA network, is determined by the ratio of the sum of OCIF propagation gain to the own-cell signal propagation gain.

\subsection {OFDMA system}
Assume that the BS transmission power is equally distributed among all sub-carriers whilst each subcarrier is assigned to only one user which has the best cannel gain \cite{Jiho2003}. $\gamma^{*}$ denotes the SINR target per subcarrier $s$ instead of per mobile $u$. The following can be obtained:
%
%
%
%
  \begin{align}\label{eqn:SINROFDMA}
                \gamma^{*}=\frac{1}{f_s+\frac{N_s\sigma^2_s}{P_b}h_s}
               \end{align}
where
\begin{align}\label{eqn:fs}
f_s=\frac{\sum_{i\neq s}G_{i,s}}{G_{b,s}}
\end{align}
\begin{align}\label{eqn:hs}
h_s=\frac{1}{G_{b,s}}=\frac{r_b^{\eta}e^{a\xi}}{k_0r_0^{\eta}}
\end{align}
and $N_s$ denotes the total number of sub-carriers and $\sigma^2_{s}$ the AWGN power per subcarrier.

Note $f_s$ and $h_s$ obtained per subcarrier for an OFDMA system are equivalent to $f_u$ and $h_u$ of a CDMA system respectively, albeit own-cell power in IOPR of a CDMA system is reduced to the signal power per subcarrier in an OFDMA system, hence $f_s$ gives the inverse of SIR.


\section{Cell assignment by signal strength}

\begin{figure}
  \centering
  \includegraphics[width=1\linewidth]{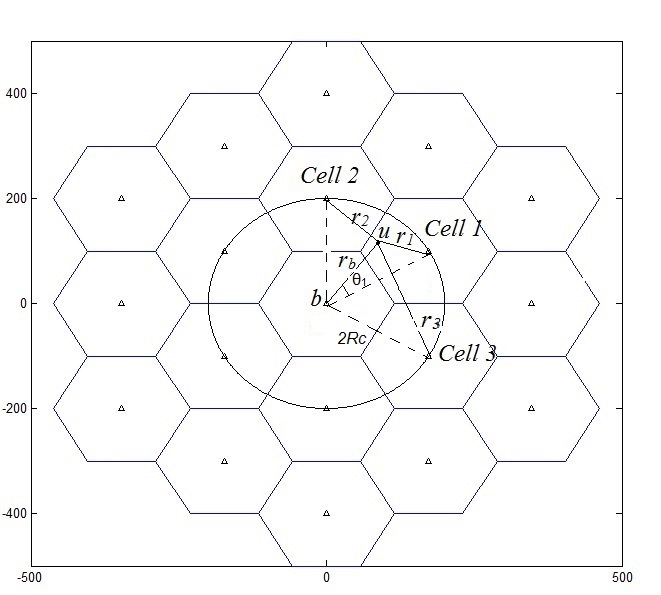}
  \caption{Grid model of hexagon cells}
  \label{geometry}
\end{figure}

Consider an area centred around BS $b$ with radius up to $2R_c$ over which mobiles are scattered uniformly, thus $r_b$ follows a probability density function $2r_b/(2R_c)^2$. When mobile $u$ is served by BS $b$, it is noted $u\subset\Psi(b)$. See Fig.\ref{geometry}. Since a mobile is attached to station $b$ from which it receives the strongest signal, we have
  \begin{align}\label{eqn:b}
  {r_b}^{-\eta}e^{-a\xi_b}>{r_j}^{-\eta}e^{-a\xi_j}\quad (for\,all\, j\neq b)
  \end{align}
where $r_j$ is the distance between the mobile and base stations other than $b$ whilst $\xi_j$ is the corresponding shadowing factor. The following can be obtained:

\begin{align}\label{eqn:CellCom}
    \xi_j-\xi_b >\frac{\eta}{a}\ln\frac{r_b}{r_j} \quad ( for\,all\,j\neq b)
\end{align}
Denote $\xi_\triangle=\xi_j-\xi_b\sim  p(\xi_\triangle)=N(0,2\sigma^{2})$. The marginal probability of a mobile attached to cell $b$ rather than cell $j$ can be obtained by:

\begin{align}\label{eqn:CellAssignment}
P_{u\not\subset\Psi(j)}=\int_{\frac{\eta}{a}\ln\frac{r_b}{r_j}}^{\infty}p{(\xi_\triangle)}\mathrm{d}\xi_\triangle=Q(\frac{\eta}{\sqrt{2}a\sigma}ln\frac{r_b}{r_j}) 
\end{align}


\begin{figure}
\centering
  \includegraphics[width=1\linewidth]{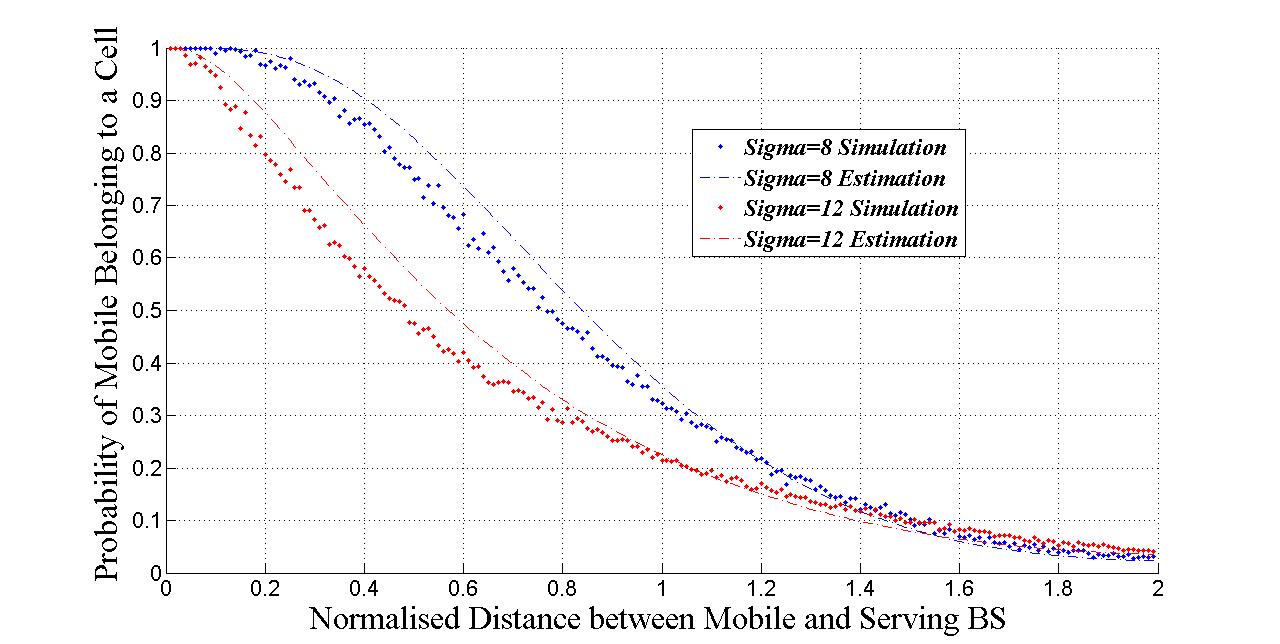}
   \caption{The probability of $P_{u\subset\Psi(b)}$ for $\eta=3$}\label{prob}
\end{figure}

Equation (\ref{eqn:CellAssignment}) shows that the larger the distance, the less likely that a mobile would attach to a BS due to path-loss. Without losing generality, the probability of $u\subset\Psi(b)$ determined by the product of the marginal probabilities of a mobile not belonging to any co-channel cell but $b$ can be approximated by that of a number of ``nearest cells" of $b$ (depending the deviation of shadowing):

\begin{align}
P_{u\subset\Psi(b)}=\prod_{j\neq b} P_{u\not\subset\Psi(j)}\approx \prod_{j\subset\{BS_\text{near}\}} P_{u\not\subset\Psi(j)}
\end{align}

Note that $P_{u\subset\Psi(b)}$ is a function of the ratio $r_b/R_c$ irrespective of the cell size $R_c$. Simulations of cell selection among six tiers of co-channel cells (total 37 cells) for a grid model network of frequency reuse factor one confirm that, the cell assignment can be determined by two smallest marginal probabilities for LN shadowing between $8dB$ and $10dB$, and by three smallest marginal probabilities for LN shadowing between $10dB$ and $12dB$. This is shown in Fig.\ref{prob}, for which a closed form estimation of the marginal probability of a mobile attached to cell $b$ rather than three other nearest cells is obtained by:
 \begin{align}
{P}_{u\not\subset\Psi(j)}\approx Q(\frac{\eta}{\sqrt{2}a\sigma}\ln\frac{r_b}{\bar{r}_j}) \quad (j=1,2,3)
\end{align}
where
\begin{equation}
 \begin{split}
 &\bar{r}_1\approx\sqrt{(r_b)^2+(2R_c)^2-4R_cr_b\cos(\pi/12)} \\
 &\bar{r}_2\approx\sqrt{(r_b)^2+(2R_c)^2-4R_cr_b\cos(\pi/3-\pi/12)}\\
 &\bar{r}_3\approx\sqrt{(r_b)^2+(2R_c)^2-4R_cr_b\cos(\pi/3+\pi/12)}
\end{split}
 \end{equation}
where the angle opposite to the $r_j$ in the acute triangle of base station $b$, $j$ and mobile $u$, that has a small range of radian values (less or equal to $\pi/6$, see Fig.\ref{geometry}), is approximated by the median.

It can also be seen from Fig.\ref{prob} that the probability of a mobile attachment decreases with increasing distance to the BS, nonetheless a mobile may distribute outside the hexagon border of a cell to which it belongs. Due to cell assignment by signal strength, mobiles that are attached to cell $b$ have a non-uniform distribution that can be obtained by:
\begin{align}
p(r_b)=\frac{r_b}{2{R_c}^2}P_{u\subset\Psi(b)}
\end{align}

\subsection {Own-cell power}
A conjecture derived from equation (\ref{eqn:CellCom}) is that, as a result of best-cell assignment, the distribution of own-cell power is truncated at small values such that
\begin{align}
P_{u\subset\Psi(b)}=\int_{-\infty}^{\xi_{b\text{max}}}p(\xi_b)\mathrm{d}\xi_b
\end{align}
Accordingly
\begin{align}
\xi_{b\text{max}}=Q^{-1}(1-P_{u\subset\Psi(b)})\sigma
\end{align}
Hence the average propagation gain of owncell signal can be obtained by
\begin{align}
\begin{split}
       &{\bar{G}_{b,u}}(r_b)=\frac{k_0r_0^{\eta}r_b^{-\eta}}{P_{u\subset\Psi(b)}}\int_{-\infty}^{\xi_{b\text{max}}} e^{(-a\xi_b)}p(\xi_b)\mathrm{d}\xi_b\\
       &=\frac{k_0r_0^{\eta}r_b^{-\eta}e^{\frac{\sigma^2a^2}{2}}}{P_{u\subset\Psi(b)}}\left(1-Q\left(Q^{-1}\left(1-P_{u\subset\psi(b)}\right)+a\sigma\right)\right)\\
\end{split}
\end{align}
This is verified by simulation. Note the result would be the same for signal gain per subcarrier $G_{b,s}$ but with different frequency related factor $k_0$. The truncation effect on the distribution of shadowing factor $\xi_b$ will be exploited in the next section.

\section {Spatial distribution of OCIF and IOPR }
\subsection{OCIF}

It is easy to understand that shadowing effect may be significant for the OCIF from near BSs whilst path-loss may outweigh shadowing in OCIF from far BSs. In the following, the average of OCIF propagation gain is derived for the sum of co-channel interference from the two nearest cells and cells beyond:

 \begin{align}\label{eqn:G}
    \sum_{i\neq b}\bar{G}_{i,u}(r_b)=\bar{G}_{1,u}(r_b)+\bar{G}_{2,u}(r_b)+\bar{G}_{3+,u}(r_b)
\end{align}
where $\bar{G}_{1,u}$ and $\bar{G}_{2,u}$ correspond to the average propagation gain of the interference from the two nearest co-channel BSs respectively whilst $\bar{G}_{3,u}$  accounts for that of the third nearest co-channel BS and beyond. The path-loss and shadowing effects are treated independently, the following can be obtained:
\begin{align}
 \begin{split}
    &\bar{G}_{j,u}(r_b)=\frac{k_0r_0^{\eta}\bar{r}_j^{-\eta}}{P_{u\subset\Psi(b)}}\int_{-\infty}^{\xi_{b\text{max}}}
    \int_{\frac{\eta}{a}\ln\frac{r_b}{\bar{r}_j}+\xi_b}^{\infty}e^{-a\xi_j}p{(\xi_j)}p{(\xi_b)}\mathrm{d}\xi_j\mathrm{d}\xi_b\\
    &=\frac{k_0r_0^{\eta}\bar{r}_j^{-\eta}e^\frac{\sigma^{2}a^2}{2}}{P_{u\subset\Psi(b)}}\int_{-\infty}^{\xi_{b\text{max}}}Q\left(\frac{\eta}{a\sigma}\ln\frac{r_b}{\bar{r}_j}+\frac{\xi_b}{\sigma}+a\sigma\right)p(\xi_b)\mathrm{d}\xi_b\\
    & \hspace {60mm} (j=1,2)
     \end{split}
      \end{align}
and
\begin{align}
\begin{split}
&\bar{G}_{3+,u}(r_b)\geq\frac{2\pi\rho_{BS}k_0r_0^{\eta}}{P_{u\subset\Psi(b)}}\int_{\bar{r}_d}^{r_{\infty}}r^{1-\eta}\\
&\hspace{20mm}\int_{-\infty}^{\xi_{b\text{max}}}\int_{\frac{\eta}{a}\ln\frac{r_b}{\bar{r_3}}+\xi_b}^{\infty}e^{-a\xi}p{(\xi)}\mathrm{d}r\mathrm{d}\xi\mathrm{d}\xi_b\\
&=\frac{2\pi\rho_{BS}k_0r_0^{\eta}}{P_{u\subset\Psi(b)}(\eta-2)}(\bar{r}_d^{2-\eta}-r_{\infty}^{2-\eta})e^\frac{\sigma^2a^2}{2}\\
&\hspace{20mm}\int_{-\infty}^{\xi_{b\text{max}}}Q\left(\frac{\eta}{a\sigma}\ln\frac{r_b}{\bar{r}_3}+\frac{\xi_b}{\sigma}+a\sigma\right)p(\xi_b)\mathrm{d}\xi_b\\
\end{split}
\end{align}
 whereby
\begin{align}
    \bar{r}_d=(\bar{r}_2+\bar{r}_3)/2
\end{align}
and $\rho_{BS}=\frac{1}{2\sqrt{3}Rc^2}$ is the base station density. The integration over a circular ring area centered around $u$ between an inner radius of $\bar{r}_d$ and outer radius of $\bar{r}_{\infty}$ can be referenced to the fluid OCIF model in \cite{Kelif2009CellBreathing}.

\begin{figure}[htbp]
   \centering
  \includegraphics[width=1\linewidth]{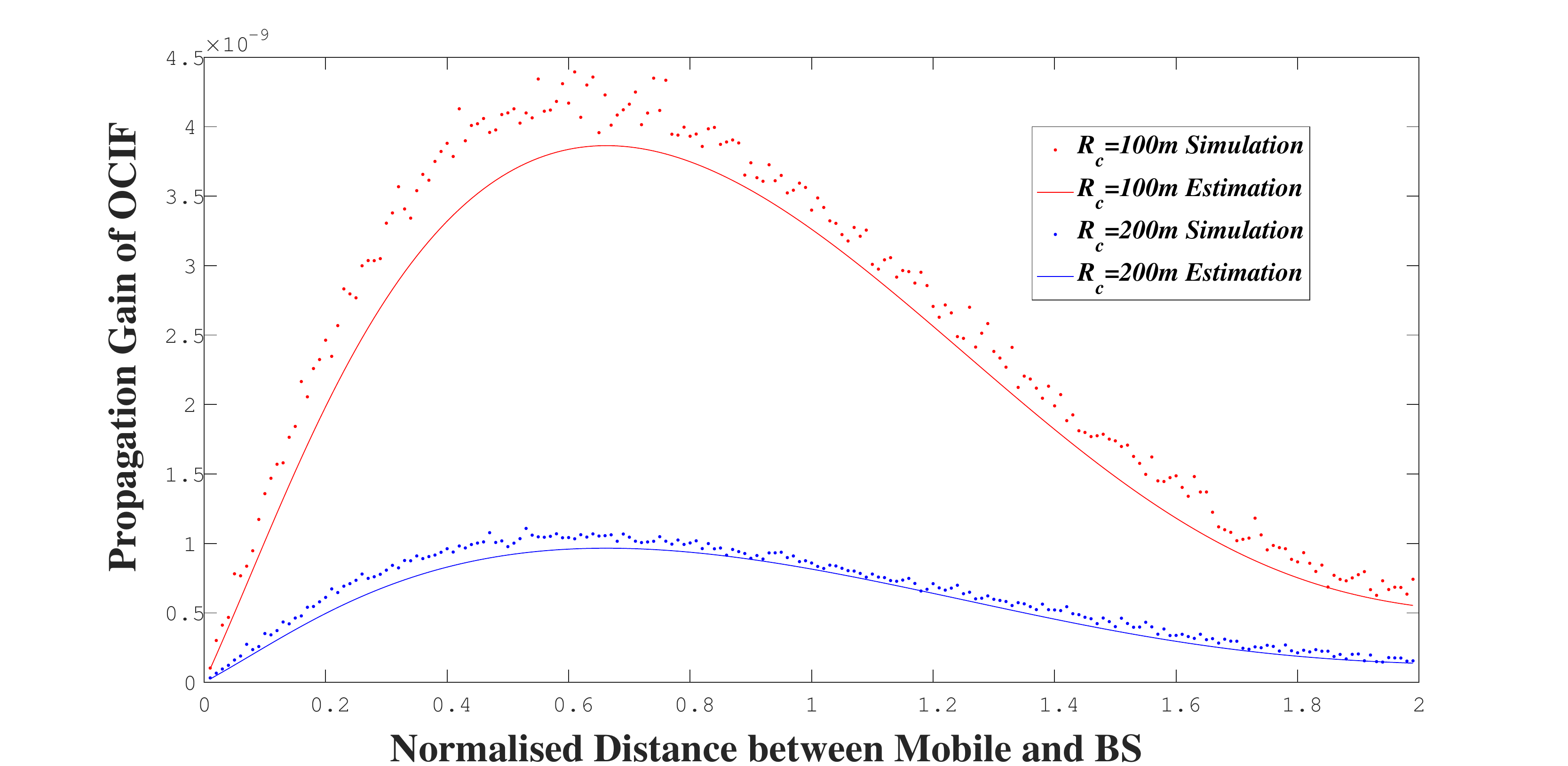}
  \caption{The Distribution of OCIF ($\eta=3$ and $\sigma=8dB)$}\label{OCIF@sigma8dB}
\end{figure}

\begin{figure}[htbp]
   \centering
  \includegraphics[width=1\linewidth]{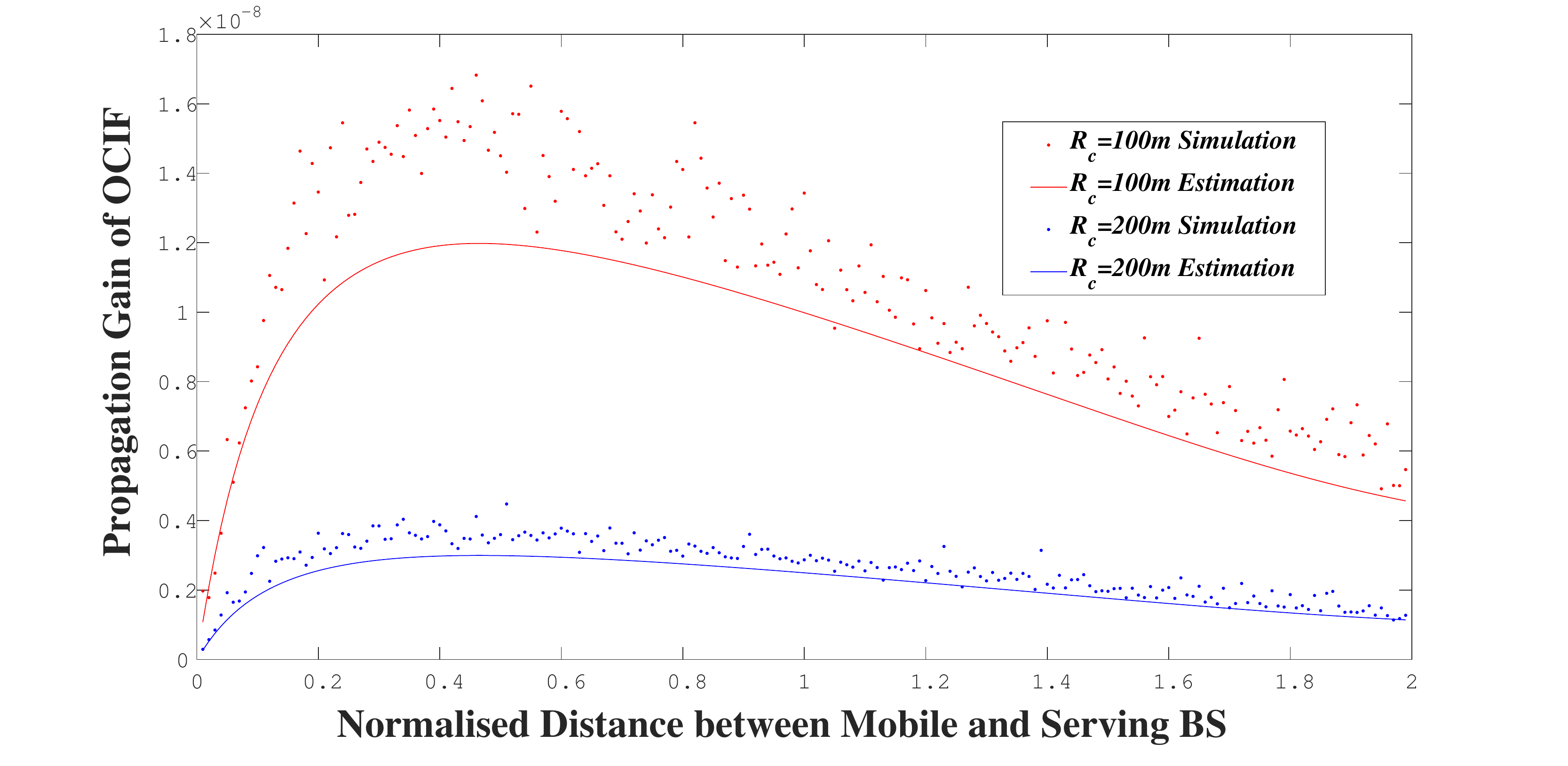}
  \caption{The Distribution of OCIF ($\eta=3$ and $\sigma=12dB)$}\label{OCIF@sigma12dB}
\end{figure}

Fig.\ref{OCIF@sigma8dB} and Fig.\ref{OCIF@sigma12dB} demonstrate that the spatial distribution of OCIF propagation gain over the distance between the mobile and its serving BS in terms of
\begin{align}
\frac{p(r_b)\sum_{i\neq b}G_{i,u}(r_b)}{\int_{0_+}^{2R_{c-}}p(r_b)\mathrm{d}r_b}
\end{align}
It can be seen that the estimation provides a tight lower bound for OCIF propagation gain, especially for a smaller shadowing effect. As anticipated, the larger the cell size, the smaller the OCIF; on the other hand, the bigger the shadowing effect, the larger the OCIF.

The spatial mean of OCIF can be obtained by:

\begin{align}\label{eqn:MuG}
\mu_G=\frac{\int_{0_+}^{2R_{c-}}p(r_b)\sum_{i\neq b}G_{i,u}(r_b)\mathrm{d}r_b}{\int_{0_+}^{2R_{c-}}p(r_b)\mathrm{d}r_b}
\end{align}
Again the result would be the same for OCIF gain per subcarrier $\sum_{i\neq b}G_{i,s}$ but with a different frequency related factor $k_0$.

\subsection{IOPR}
\begin{figure}[htbp]
  \centering
  \includegraphics[width=1\linewidth]{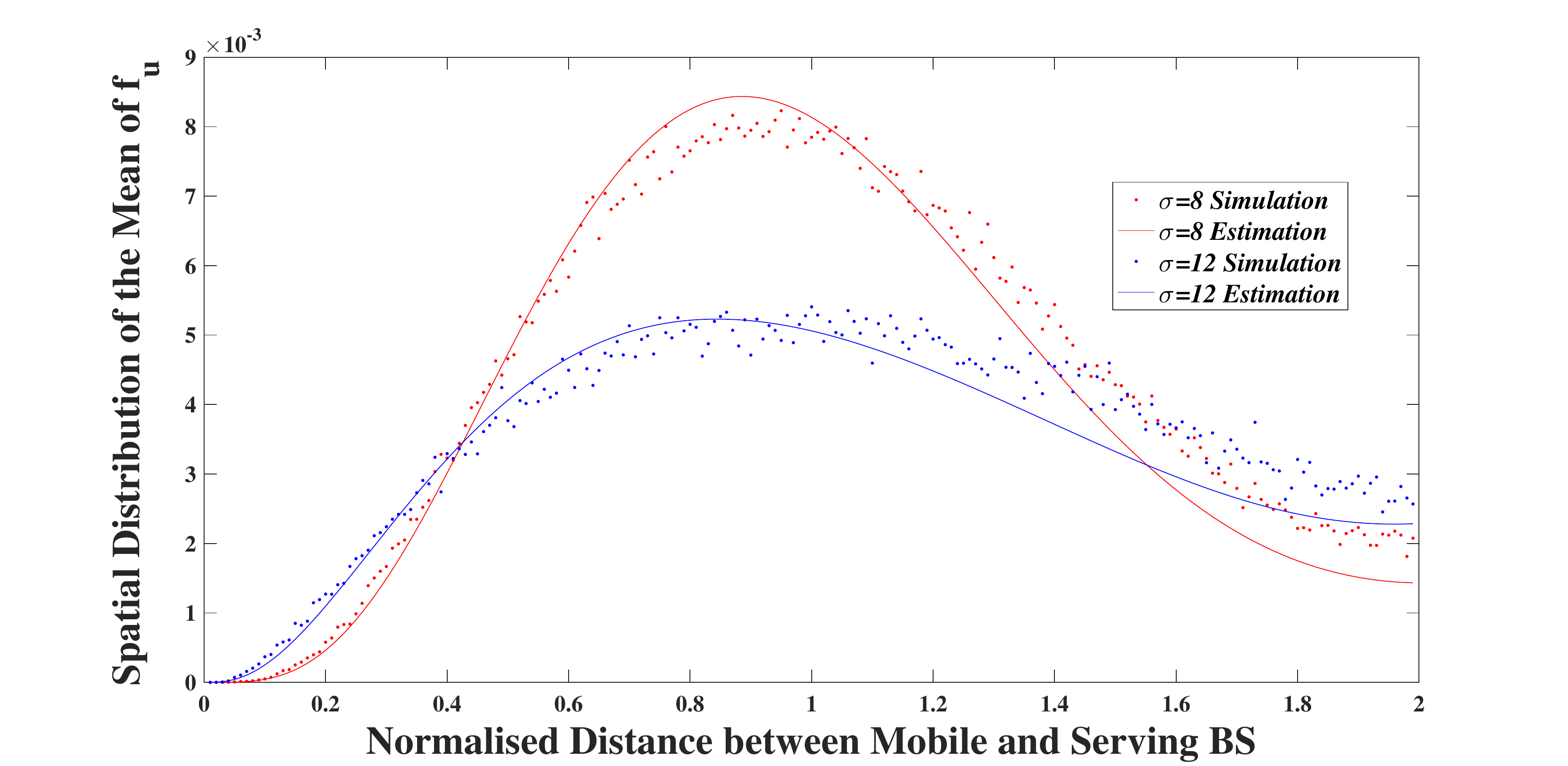}
  \caption{The distribution of $\bar{f}$ for $\eta=3$ }\label{fu}
\end{figure}

\begin{figure}[htbp]
  \centering
  \includegraphics[width=1\linewidth]{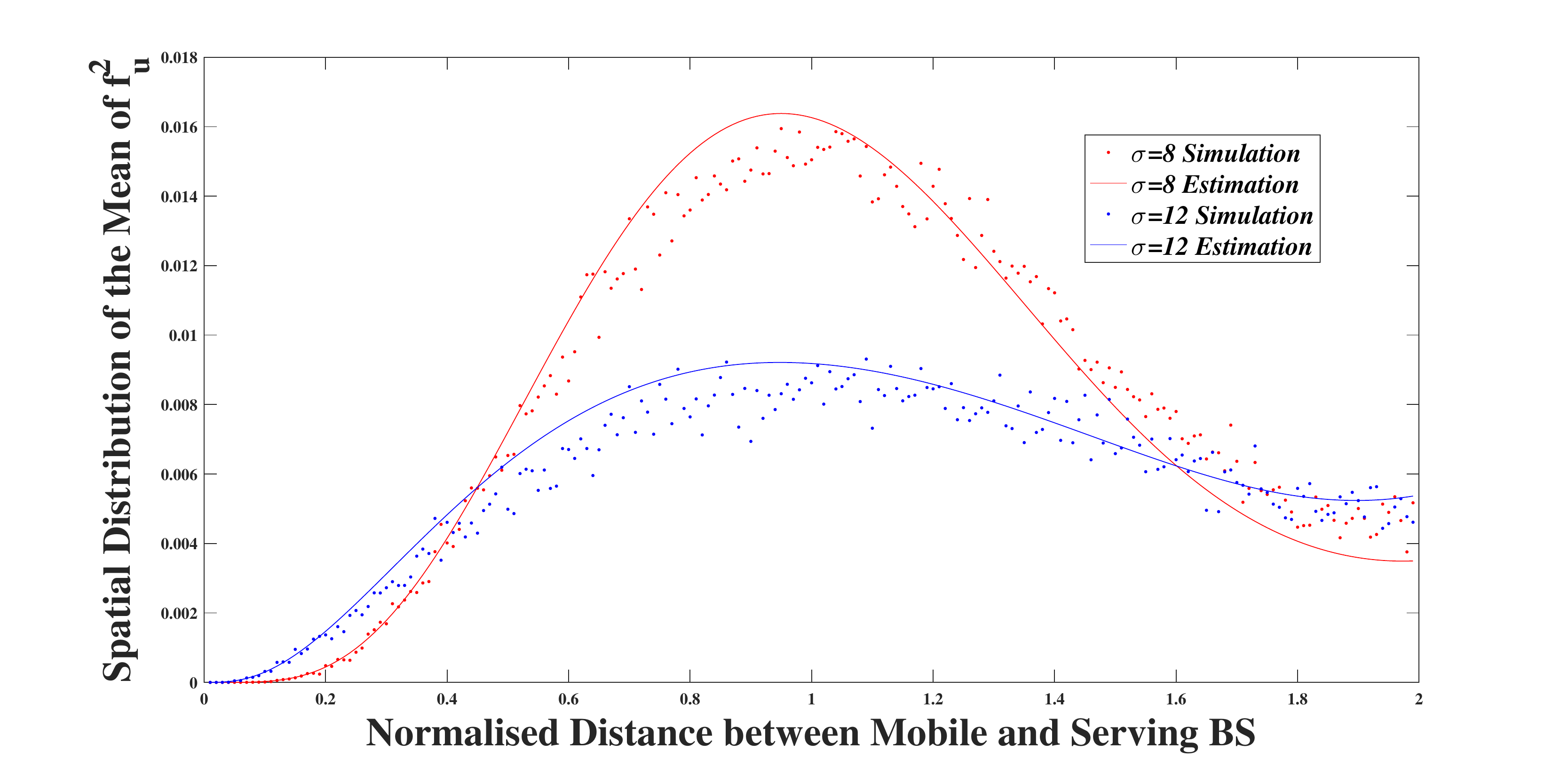}
  \caption{The distribution of $\bar{f^2}$ for $\eta=3$ }\label{fu^2}
\end{figure}

Likewise, the average of either $f_u$ given in equation (\ref{eqn:fu}) or $f_s$ in equation (\ref{eqn:fs}) can be derived for the sum of the contributions from the two nearest cells and cells beyond as in the following (the indices $u$ and $s$ are dropped for brevity):
\begin{align}
    \bar{f}(r_b)=\bar{f}_1(r_b)+\bar{f}_2(r_b)+\bar{f}_{3+}(r_b)
\end{align}
where
\begin{align}
 \begin{split}
    &\bar{f}_j(r_b)=\frac{1}{P_{u\not\subset\Psi(j)}}{(\frac{r_b}{\bar{r}_j})}^{\eta}\int_{\frac{\eta}{a}\ln\frac{r_b}{\bar{r}_j}}^{\infty}e^{-a\xi_\triangle}p{(\xi_\triangle)}\mathrm{d}\xi_\triangle\\
    &=\frac{e^{\sigma^{2}a^2}}{P_{u\not\subset\Psi(j)}}{(\frac{r_b}{\bar{r}_j})}^{\eta}Q\left(\frac{\eta}{\sqrt{2}a\sigma}\ln\frac{r_b}{\bar{r}_j}+\sqrt{2}a\sigma+\delta_j\right)\\
     & \hspace {55mm} (j=1,2)
     \end{split}
      \end{align}
and
\begin{align}
\begin{split}
&\bar{f}_{3+}(r_b)\geq\frac{2\pi\rho_{BS}r_b^{\eta}}{P_{u\not\subset\Psi(3)}}\int_{\bar{r}_d}^{r_{\infty}}r^{1-\eta}\mathrm{d}r\int_{\frac{\eta}{a}\ln\frac{r_b}{\bar{r}_3}}^{\infty}e^{-a\xi_\triangle}p{(\xi_\triangle)}\mathrm{d}\xi_\triangle\\
&=\frac{2\pi\rho_{BS}r_b^{\eta}}{P_{u\not\subset\Psi(3)}(\eta-2)}(\bar{r}_d^{2-\eta}-r_{\infty}^{2-\eta})e^{\sigma^{2}a^2}\\
&\hspace{30mm} Q\left(\frac{\eta}{\sqrt{2}a\sigma}\ln\frac{r_b}{\bar{r}_3}+\sqrt{2}a\sigma+\delta_3\right)\\
\end{split}
\end{align}
in which
  \begin{align}
  \delta_j=\left(Q^{-1}(P_{u\subset\Psi(b)})-\frac{\eta}{\sqrt{2}a\sigma}\ln\frac{r_b}{\bar{r}_j}\right)/\sigma \hspace {5mm} (j=1,2,3)
  \end{align}
accounts for the shift of the mean of $\xi_\triangle$ from a marginal probability $u\not\subset\Psi(j)$ to $u\subset\Psi(b)$. Similarly we have
\begin{align}
 \begin{split}
    &\overline{f_j^2}(r_b)=\frac{1}{P_{u\not\subset\Psi(j)}}{(\frac{r_b}{\bar{r}_j})}^{2\eta}\int_{\frac{\eta}{a}\ln\frac{r_b}{\bar{r}_j}}^{\infty}e^{-2a\xi_\triangle}p{(\xi_\triangle)}\mathrm{d}\xi_\triangle\\
    &=\frac{e^{4\sigma^{2}a^2}}{P_{u\not\subset\Psi(j)}}{(\frac{r_b}{\bar{r}_j})}^{2\eta}Q\left(\frac{\eta}{\sqrt{2}a\sigma}\ln\frac{r_b}{\bar{r}_j}+2\sqrt{2}a\sigma+\delta_j\right)\\
     & \hspace {55mm} (j=1,2)
     \end{split}
      \end{align}
and
\begin{align}
\begin{split}
&\overline{f_{3+}^2}(r_b)\geq\frac{\left[2\pi\rho_{BS}\int_{\bar{r}_d}^{r_{\infty}}{(\frac{r}{r_b})}^{-\eta}r\mathrm{d}r\right]^2}{P_{u\not\subset\Psi(3)}}\int_{\frac{\eta}{a}\ln\frac{r_b}{\bar{r}_3}}^{\infty}e^{-2a\xi_\triangle}p{(\xi_\triangle)}\mathrm{d}\xi_\triangle\\
&=\frac{\left[\frac{2\pi\rho_{BS}r_b^{\eta}}{(\eta-2)}(\bar{r}_d^{(2-\eta)}-r_{\infty}^{(2-\eta)})\right]^2}{P_{u\not\subset\Psi(3)}}\\
&\hspace{30mm}e^{4\sigma^{2}a^2}Q\left(\frac{\eta}{\sqrt{2}a\sigma}\ln\frac{r_b}{\bar{r}_3}+2\sqrt{2}a\sigma+\delta_3\right)
\end{split}
\end{align}
Fig.\ref{fu}  and Fig.\ref{fu^2} show that the distributions of both first-order and second-order of $f_u$ estimated respectively by:
 \begin{align}
\frac{\overline{f}(r_b) p(r_b)}{\int_{0_+}^{2R_{c-}}p(r_b)\mathrm{d}r_b}
\end{align}
and
\begin{align}
\frac{\overline{f^2}(r_b) p(r_b)}{\int_{0_+}^{2R_{c-}}p(r_b)\mathrm{d}r_b}
\end{align}
give good estimations. It is easy to understand that the distributions are getting flatter for a larger shadowing effect due to best cell configuration. The spatial mean and variance of $f$ can be calculated below:

\begin{align}
\mu_f=\frac{\int_{0_+}^{2R_{c-}}\bar{f}(r_b)p(r_b)\mathrm{d}r_b}{\int_{0_+}^{2R_{c-}}p(r_b)\mathrm{d}r_b}\
\end{align}
\begin{align}
\sigma_f^2=\frac{\int_{0_+}^{2R_{c-}}\overline{f^2}(r_b)p(r_b)\mathrm{d}r_b}{\int_{0_+}^{2R_{c-}}p(r_b)\mathrm{d}r_b}-\mu_f^2
 \end{align}
where
\begin{align}
\begin{split}
&\overline{f^2}(r_b)=\overline{f_1^2}(r_b)+\overline{f_2^2}(r_b)+\overline{f_{3+}^2}(r_b)\\
&+2\bar{f}_1(r_b)\bar{f}_2(r_b)+2\bar{f}_1(r_b)\bar{f}_3(r_b)+2\bar{f}_2(r_b)\bar{f}_3(r_b)\\
\end{split}
\end{align}
Note both $\mu_f$ and $\sigma_f$ are independent of the absolute vale of the cell size $R_c$. This is because the own-cell power and other cell power cancel each other in a homogenous network.

\section{Network Dimensioning: an OFDMA system }

With dynamic dimensioning, coverage for a varying traffic demand needs to be satisfied with minimum number of sites and minimum power cost. On one hand, the coverage defined by the probability that a random user can achieve a target SINR (indication of maximum rate by Shannon theorem) can be obtained as the average of complimentary cumulative  distribution  function (CCDF) of SINR over the entire area of coverage \cite{andrews2011}. On the other hand, outage, the complement of coverage, is the probability that the SINR threshold cannot be met for a random user attached to the cell \cite{Kelif2009CellBreathing}. Noise power noise power spectrum density $N_0=4*10^{-21}$ Watts/Hz, and system total bandwidth $B_w=5$ MHz are assumed in the following subsections.

\subsection {BS transmission power per cell size}

\begin{figure}[htbp]
  \centering
  \includegraphics[width=1\linewidth]{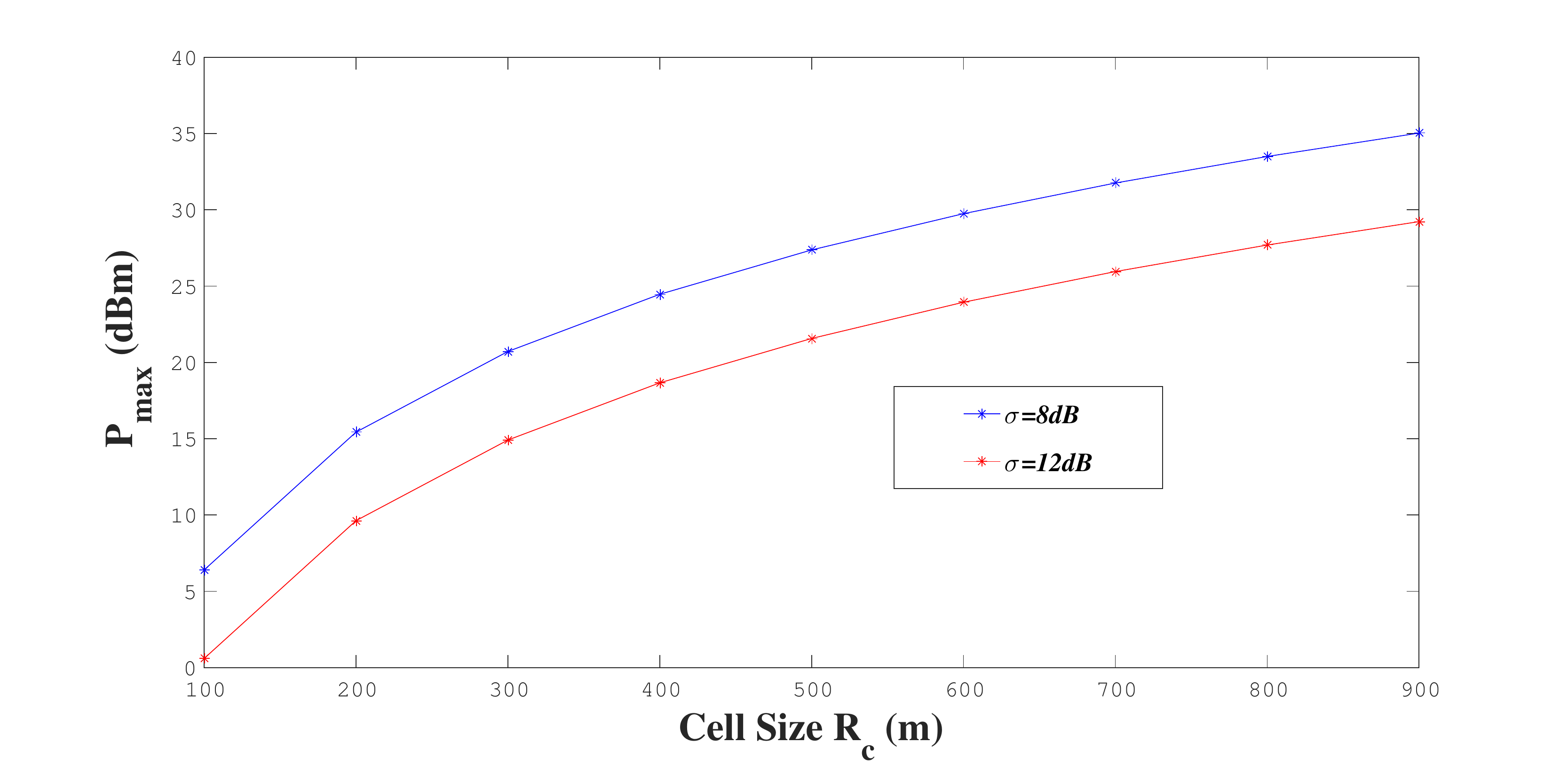}
  \caption{Ideal $P_{max}$ for different cell size for $\eta=3$}\label{Pmax}
\end{figure}

\begin{figure}[htbp]
 \centering
  \includegraphics[width=1\linewidth]{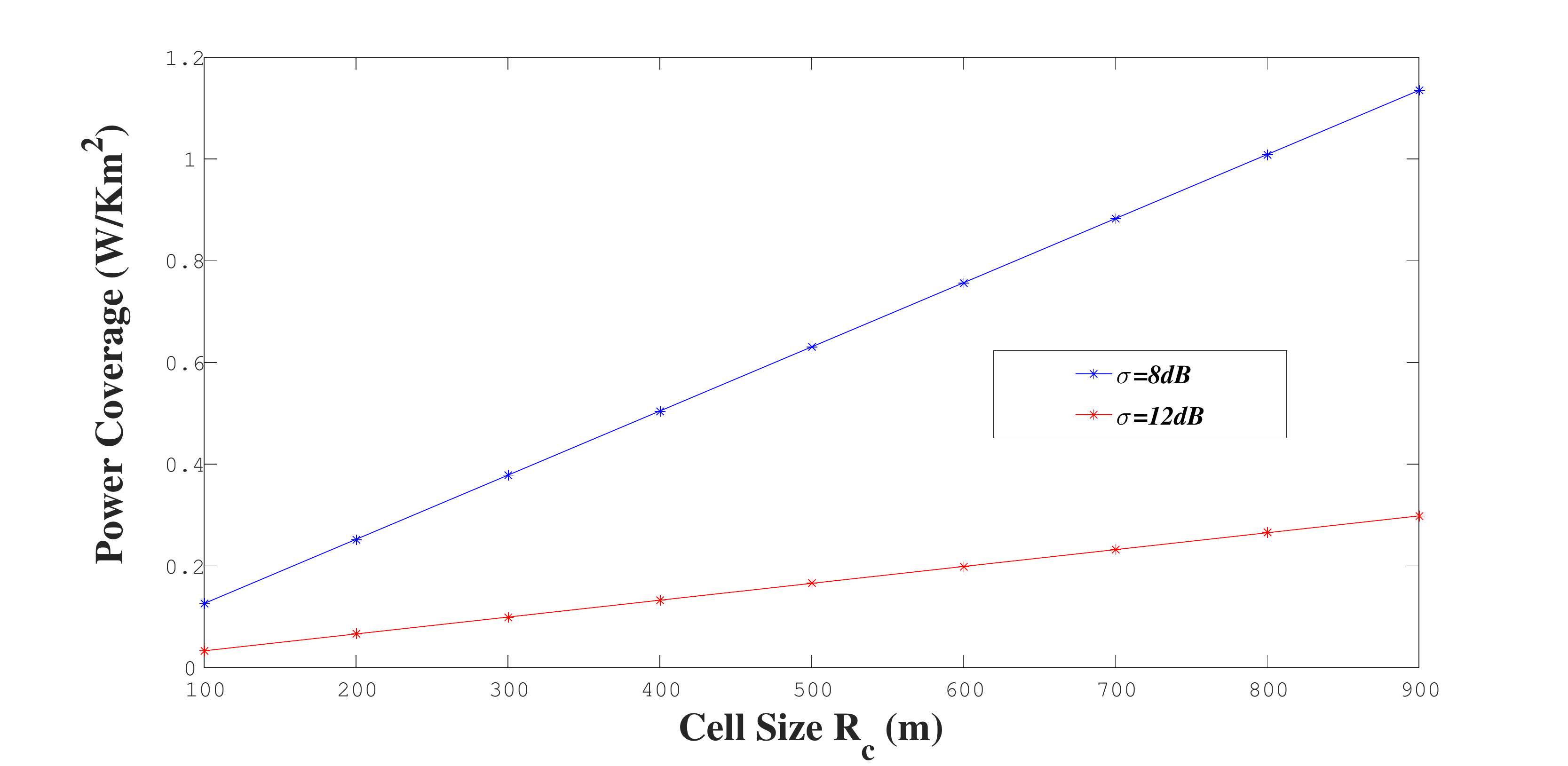}
 \caption{Power coverage over area vs cell size for $\eta=3$}\label{PowerD}
\end{figure}

Referring to equation (\ref{eqn:SINROFDMA}), the following has to be satisfied to guarantee an interference-limited system:
\begin{align}
 f_s\gg\frac{N_s\sigma_s^2}{P_b}h_s
 \end{align}
of which $f_s$ is the inverse of SINR.  Assume a large ratio of $10^5:1$ between two sides of the above inequality, the maximum BS power can be obtained as:
\begin{align}
       P_{max}=\frac{10^5N_s\sigma_s^2h_s}{f_s}=\frac{10^5N_s\sigma_s^2}{\sum_{i\neq b}G_{i,s}}
\end{align}
where $N_s\sigma_s^2=N_0B_w$, gives the total noise power of an equivalent broadband system. Replace $\sum_{i\neq b}G_{i,s}$ with its mean in equation (\ref{eqn:MuG}), maximum base station transmission power obtained for a homogenous network are shown in Fig.\ref{Pmax} for different cell sizes. Consequently power density over area is increased with increasing cell size as shown in Fig. \ref{PowerD}

Points need to be noted. First, BS transmission power is only a small part of an overall power cost of a BS \cite{arnold2010power}. Second, the maximum BS transmission powers appear to be in line with that of a commercial BS of macro-cell, micro-cell or pico-cell, however this modeling suggests a significant reduction is possible without compromising the system being interference limited.

\subsection {Cell density to guarantee a satisfactory coverage}

The outage probability of can be deduced from (\ref{eqn:SINROFDMA}) for an interference limited system as:
\begin{align}
        P_{out}(\gamma^*)=Pr\left[f_s>\frac{1}{\gamma^*}\right]
 \end{align}
With best-cell configuration, $f_s$ is a sum of LN RVs (random variables) with larger attenuation (due to path-loss from far BSs) and truncated LN RVs  with smaller attenuation (due to path-loss from near BSs). Empirical studies suggest that $f_s$ can be approximated by a log-normal distribution. With mean $m=\bar{f}(rb)$ and variance $v=\overline{f^2}(r_b)-(\bar{f}(r_b))^2$, the outage probability can be obtained:
\begin{align}
         O(r_b)\approx Q\left(\frac{-\ln(\gamma^{*})-\mu}{\sigma}\right)\ %
\end{align}
where
\begin{align}
\mu=\ln\left(\frac{m}{\sqrt{1+v/m^2}}\right)
\end{align}
\begin{align}
\sigma=\sqrt{\ln(1+v/m^2)}
\end{align}
This is proved to be a good approximation when $\sqrt2\sigma$ is used instead of $\sigma$ to compensate for the truncation effect. The average outage probability over the entire cell is computed by:
\begin{align}\label{eqn:OutageOFDMA}
P_{out}(\gamma^*)=\frac{\int_{0_+}^{2R_{c-}}O(r_b)p(r_b)\mathrm{d}r_b}{\int_{0_+}^{2R_{c-}}p(r_b)\mathrm{d}r_b}\
\end{align}

\begin{figure}[htbp]
  \centering
  \includegraphics[width=1\linewidth]{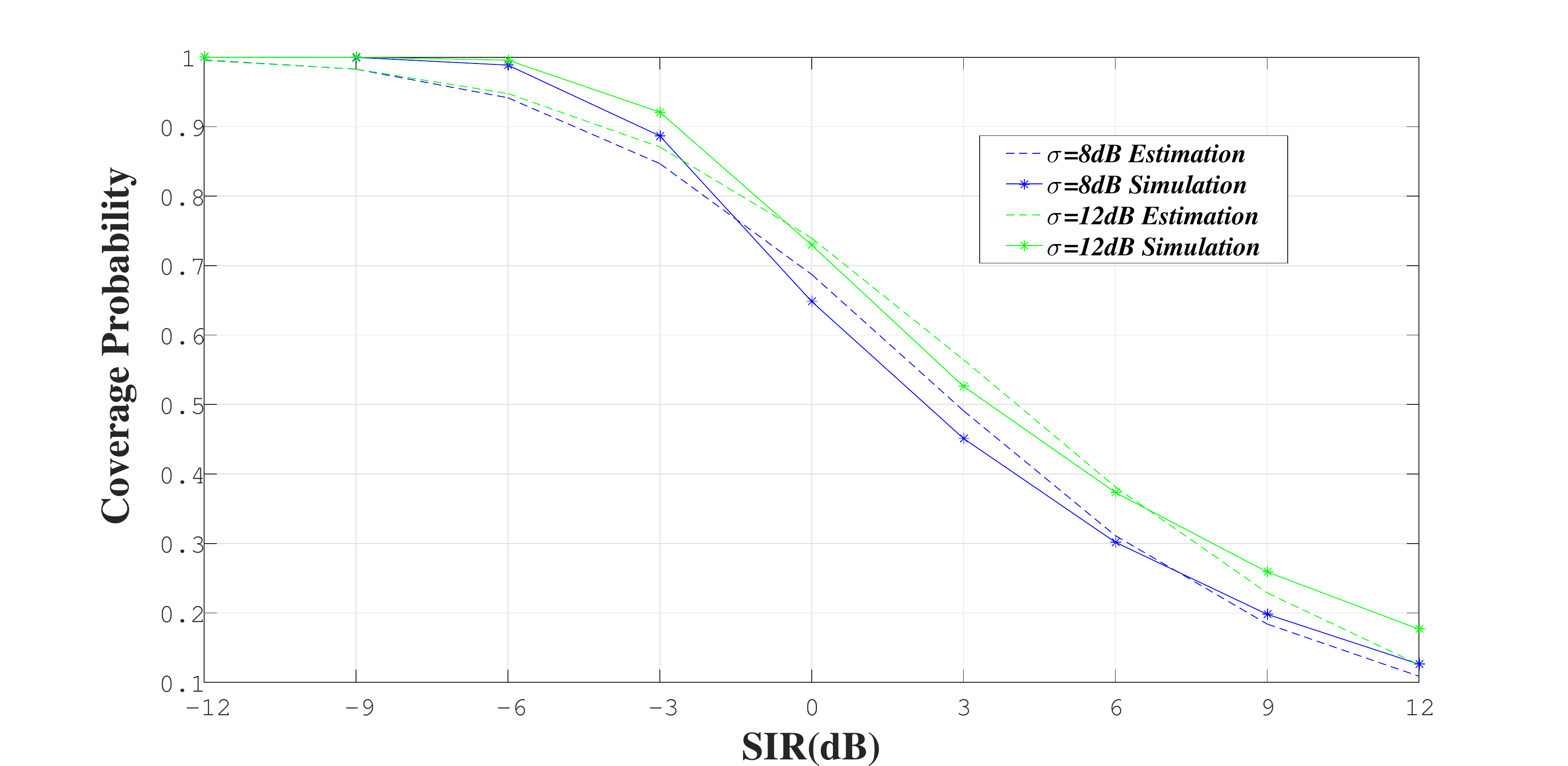}
    \caption{Probability of coverage vs. SIR for $\eta=3$}\label{OFDMACoverage}
\end{figure}

Fig.\ref{OFDMACoverage} illustrates the coverage probability vs. SIR for an interference limited OFDMA system. The analytical modeling is verified by the simulation results, which show that the coverage is actually improved slightly for a larger LN shadowing. Coinciding with the observations of an improved QoS in \cite{Bartlomiej2013}, this phenomenon is due to the best cell configuration with which shadowing effect may undermine path-loss effect and result in smaller IOPR (also seen in Fig.\ref{fu}), hence giving better coverage or more capacity.

\begin{figure}[htbp]
  \centering
  \includegraphics[width=1\linewidth]{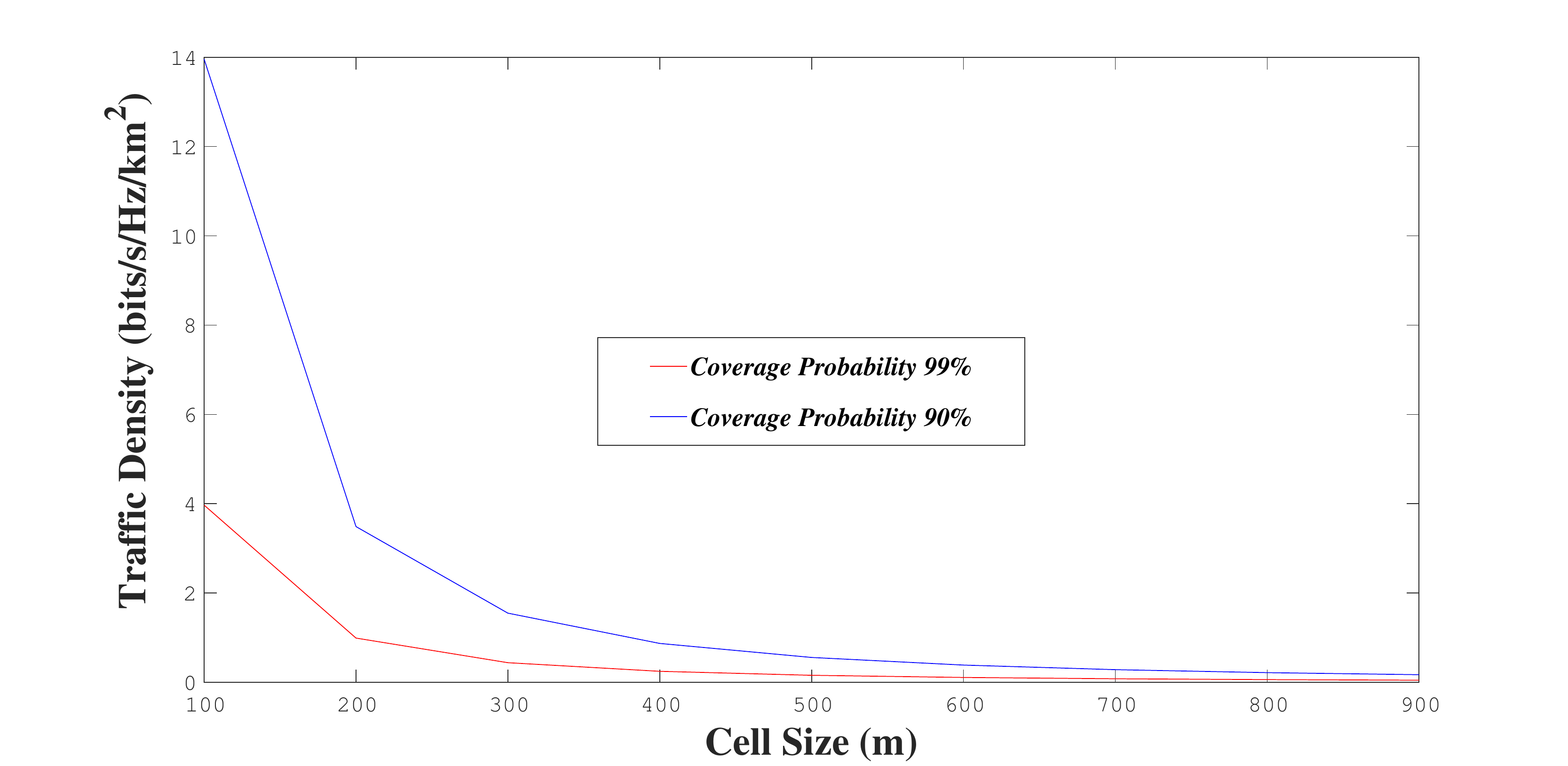}
  \caption{Rate per area for $\eta=3$ and $\sigma=8dB\sim12dB$}\label{TrafficDensityVsCellSize}
\end{figure}

Note the above results are independent of cell size. With reference to Shannon theorem:
\begin{align}\label{eqn:shannon}
C_{cell}=\log_2(1+\gamma^{*}) \quad (bits/s/Hz),
\end{align}
the trade-off relation between coverage probability and cell rate capacity can be derived jointly from (\ref{eqn:OutageOFDMA}) and (\ref{eqn:shannon}) for each carrier irrespective of cell size. On guaranteeing a satisfactory coverage, the rate density over the area are met with a corresponding cell density, as shown in Fig.\ref {TrafficDensityVsCellSize}. Rate density can be significantly higher with MIMO, but is beyond the scope of this paper.

\section{Conclusion}

In this paper, two important parameters for network dimensioning, i.e. the spatial distributions of OCIF and IOPR are modeled for cellular systems following the best cell configuration. The modeling provides more realistic evaluation of coverage and rate that are verified by simulations. The modeling is derived for a grid model network, but it is applicable to other network models so long as the distances of a mobile to a few nearest BSs can be referenced to its distance to the serving base station. The results in the paper can be easily expanded to a cellular network of sector sites with directional antennas and with frequency re-use factor other than one by changing the co-channel BS density appropriately.

In comparision to the nearest cell configuration commonly assumed in previous publications, a few conclusions can be drawn. First, the coverage can be actually improved slightly under a larger shadowing effect, this is also supported by the observations of improved QoS in \cite{Bartlomiej2013} under larger shadowing. Second, the base station transmission power can be reduced significantly with dynamic dimensioning. Third, cell assignment by signal strength means smaller variance of both OCIF and IOPR distribution, hence steeper curves of coverage probability vs. rate capacity, this is particularly obvious for a CDMA system in which the average effect on IOPR due to multiple users sharing the same band means even smaller variance. Application to the network dimensioning of CDMA systems will be given in future works.

\addcontentsline{toc}{section}{References}


\end{document}